\documentclass[notitlepage,aps,prl,reprint,superscriptaddress,longbibliography]{revtex4-1}
\usepackage[utf8]{inputenc}

\usepackage{mathtools}
\usepackage[english]{babel}
\usepackage{hyperref}
\usepackage{color}

\newcommand{\appropto}{\mathrel{\vcenter{
  \offinterlineskip\halign{\hfil$##$\cr
    \propto\cr\noalign{\kern2pt}\sim\cr\noalign{\kern-2pt}}}}}

\DeclarePairedDelimiter\bra{\langle}{\rvert}
\DeclarePairedDelimiter\ket{\lvert}{\rangle}
\DeclareMathOperator{\sinc}{sinc}

\begin{document}
%\title{Digitalization of Continuous-Variable Quantum States}
\title{Universal unitary transfer of continuous-variable quantum states into a few qubits}
\author{Jacob Hastrup}
\email{jhast@fysik.dtu.dk}
\affiliation{Center for Macroscopic Quantum States (bigQ), Department of Physics, Technical University of Denmark, 2800 Kgs. Lyngby, Denmark}
\author{Kimin Park}
\affiliation{Department of Optics, Palacky Univeristy, 77146 Olomouc, Czech Republic}
\affiliation{Center for Macroscopic Quantum States (bigQ), Department of Physics, Technical University of Denmark, 2800 Kgs. Lyngby, Denmark}
\author{Jonatan Bohr Brask}
\affiliation{Center for Macroscopic Quantum States (bigQ), Department of Physics, Technical University of Denmark, 2800 Kgs. Lyngby, Denmark}
\author{Radim Filip}
\affiliation{Department of Optics, Palacky Univeristy, 77146 Olomouc, Czech Republic}
\author{Ulrik Lund Andersen}
\affiliation{Center for Macroscopic Quantum States (bigQ), Department of Physics, Technical University of Denmark, 2800 Kgs. Lyngby, Denmark}

% \begin{abstract}
% \textbf{We present a protocol for transferring arbitrary continuous-variable quantum states into a discrete-variable qubit register and back. Our protocol is deterministic and utilizes only a few two-mode interactions which are readily available in trapped-ion and superconducting platforms. The inevitable errors caused by transferring an infinite-dimensional state into a finite-dimensional register are suppressed exponentially with the number of qubits. Furthermore, the encoded states exhibit robustness against noise acting on the register. Our protocol has applications to quantum memories, non-Gaussian state generation, non-Gaussian state manipulation and hardware-efficient multi-qubit operations, thus providing a powerful and flexible tool for discrete-continuous hybrid quantum information processing.}
% \end{abstract}
\begin{abstract}
\textbf{We present a protocol for transferring arbitrary continuous-variable quantum states into a few discrete-variable qubits and back. The protocol is deterministic and utilizes only two-mode Rabi-type interactions which are readily available in trapped-ion and superconducting circuit platforms. The inevitable errors caused by transferring an infinite-dimensional state into a finite-dimensional register are suppressed exponentially with the number of qubits. Furthermore, the encoded states exhibit robustness against noise, such as dephasing and amplitude damping, acting on the qubits. Our protocol thus provides a powerful and flexible tool for discrete-continuous hybrid quantum systems.}
\end{abstract}
\date{\today}

\maketitle
\section{Introduction}

Quantum information processing (QIP) can be realized using both discrete variables (DV), such as the energy levels of atoms or superconducting qubits, or continuous variables (CV) \cite{weedbrook2012gaussian}, such as the quadratures of an electromagnetic field, spin ensemble or mechanical oscillator. Both types of systems have various advantages and disadvantages, depending on the particular task, application and implementation. For example, universal control of noisy many-qubit systems has become available \cite{arute2019quantum}, but truly scalable systems and break-even error correction remains to be demonstrated. On the other hand, CV QIP is highly scalable, allowing long range interactions which has been used to demonstrate entanglement of millions of modes \cite{yoshikawa2016invited} and generation of 2D cluster-states \cite{asavanant2019generation,larsen2019deterministic} with current technology. Furthermore, the infinite dimensionality of a single CV mode can be utilized for hardware-efficient single-mode error correction \cite{ofek2016extending,hu2019quantum,campagne2020quantum,de2020error} and high-dimensional operations, such as the quantum Fourier transform, can be implemented with simple, single-mode operations \cite{weedbrook2012gaussian}. However, non-Gaussian operations required for universal quantum processing and fault tolerance have proven difficult to realize in pure CV systems. 

Two of the leading platforms for quantum computing are trapped ions and superconducting circuits. These systems support both DV QIP through spin or charge qubits, as well as CV QIP through motional modes or microwave cavity modes. Furthermore, the CV and DV modes can couple, enabling CV-DV hybrid interactions. In fact, it is common to utilize this hybrid interaction to enable various operations. For example, for DV QIP, the CV modes can be used to facilitate multi-mode operations and qubit read-out \cite{bruzewicz2019trapped,krantz2019quantum}. Meanwhile, for CV QIP, the DV modes are used to enable non-Gaussian operations \cite{ofek2016extending,hu2019quantum,campagne2020quantum,de2020error} which are required for universality. Thus CV-DV hybrid interactions have proven valuable in overcoming the challenges associated with either CV or DV QIP.

Here, we add a new element to the toolbox of CV-DV hybrid operations by showing that arbitrary quantum states can be coherently and deterministically mapped between a CV mode and a collection of qubits using accessible two-mode interactions. This mapping has several potential applications for QIP. For example, our scheme enables qubit-based memories for CV states. Many types of CV QIP relies on heralded, non-deterministic operations and are therefore dependent on quantum memories. A qubit-based memory could enable DV error correction protocols to be carried out on arbitrary CV states. Additionally, if the qubits are coupled to two different CV modes, e.g. transmon qubits coupled to both a mechanical acoustic mode and a microwave cavity mode, one CV mode can be encoded to the qubits and then decoded onto the other CV mode, enabling qubit-mediated transfer of CV information from one CV mode to another. Furthermore, our scheme can also be used for efficient deterministic generation of arbitrary CV states, such as non-Gaussian states, by preparing the qubits in an equivalent encoded state and then applying the inverse mapping to transfer the state to the CV mode. In general, applications of this protocol will strongly depend on the physical system but promise to aid in solving a wide range of issues in hybrid QIP platforms. 

Unlike previous proposal for transferring CV states onto qubits \cite{fiuravsek2002encoding,chen2006quantum}, our protocol makes efficient use of the available qubit dimensionality, such that only a few qubits are required, while also using experimentally available interactions.

\section{Protocol}
\begin{figure}[t!]
    \centering
    \includegraphics{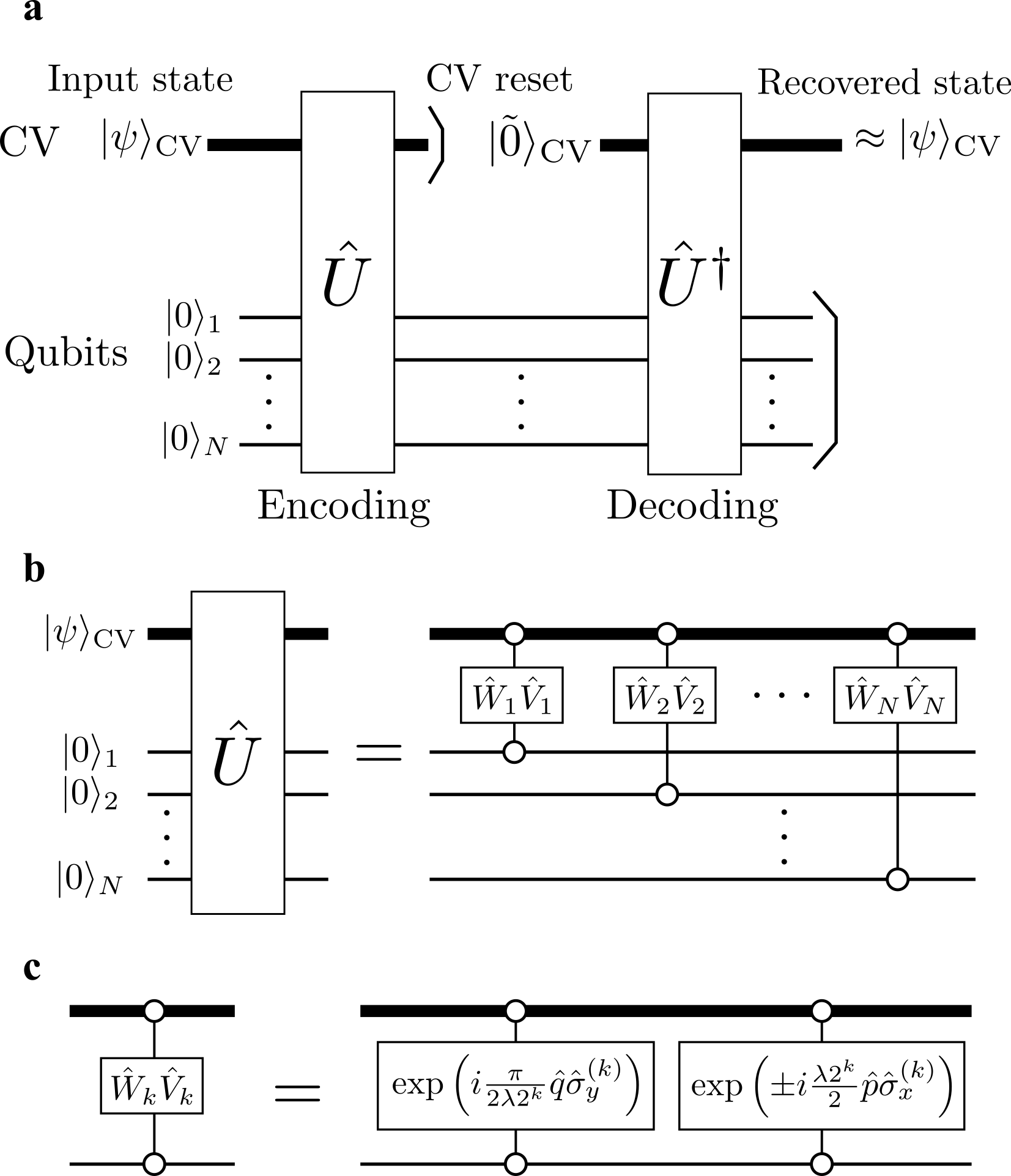} 
    \caption{(a): Circuit of encoding and decoding to transfer a CV state to a collection of qubits and back. (b): The encoding is achieved by interacting the CV mode sequentially with each of the qubits. (c): Each interaction unitary is composed of two Rabi interactions as given by equation \eqref{eq:Rabi}.}
    \label{fig:circuit}
\end{figure}
The system we are considering consists of a single CV mode and $N>1$ qubits, as illustrated in Fig.\ \ref{fig:circuit}a. The protocol is designed to transfer an arbitrary CV state $\ket{\psi}_\textrm{CV}$ into an entangled state of the qubits, leaving the CV mode in an input-independent state, which we denote $\ket{\tilde{0}}_\textrm{CV}$. Since the CV mode has an infinite dimensionality while the qubits have a finite dimension, such a protocol is in principle impossible for arbitrary states. However, in practice we can expect relevant input CV states to have majority of their support in a finite-dimensional subspace, thereby allowing a CV-DV mapping to a good approximation. Furthermore, since the dimension of the qubit subspace scales exponentially, i.e. $2^N$, with the number of qubits, $N$, we can expect the approximation to become very good with only a few qubits. In general, the protocol can be described by the following unitary operation:
\begin{equation}
\hat{U}\Big[\ket{\psi}_\textrm{CV}\ket{\textbf{0}}_\textrm{DV}\Big]=\sqrt{(1-\varepsilon)}\ket{\tilde{0}}_\textrm{CV}\ket{\Psi}_\textrm{DV} + \sqrt{\varepsilon}\ket{\Phi_\varepsilon}_\textrm{CV/DV},
\label{eq:map}
\end{equation}
where $\ket{\mathbf{0}}_\textrm{DV}=\bigotimes_{k=1}^N\ket{0}_k$ is the product of the ground states of the qubits, $\ket{\Psi}_\textrm{DV}$ is the encoded DV state and $\ket{\Phi_\varepsilon}_\textrm{CV/DV}$ is a residual entangled CV-DV state defined to make $\hat{U}$ unitary and such that $\langle\Phi_\varepsilon|\tilde{0}\rangle=0$. $\varepsilon$ is a real parameter, $0\leq \varepsilon \leq 1$, quantifying the error of the protocol, e.g. due to the CV-DV dimensionally mismatch. $\varepsilon$ thus depends on the input state, and a successful protocol should aim to minimize $\varepsilon$ for a large class of input states. For $\varepsilon = 0$ the mapping is perfect as the CV mode contains no information about the input state, i.e. the state has been perfectly transferred to the DV modes. 

The input state can be recovered by applying $\hat{U}^\dagger$. If the CV mode is completely reset to the state $\ket{\tilde{0}}_\textrm{CV}$ after the application of $\hat{U}$, the fidelity, $F$, between the input and recovered state is related to $\varepsilon$ by:
\begin{equation}
(1-\varepsilon)^2 \leq F \leq 1-\varepsilon, \label{eq:fidelity}
\end{equation}  
with the exact value of $F$ depending on the input state (details are given in Appendix A).

We now show how to decompose $\hat{U}$ into experimentally accessible two-mode interactions. A circuit diagram of the encoding unitary is shown in Fig.\ \ref{fig:circuit}b. It consists of $N$ interaction terms, each of which are composed of two interactions between the CV mode and one of the qubits:
\begin{equation}
\hat{U}=\prod_{k=1}^N \hat{W}_k \hat{V}_k. \label{eq:product}
\end{equation}
The interactions are conditional displacements \cite{haljan2005spin,fluhmann2018sequential,campagne2020quantum} which are generated by a Rabi-type Hamiltonian, i.e. a coupling between a quadrature operator of the CV mode and a Pauli operator of the qubit:
\begin{align}
\begin{split}
\hat{V}_k &= \exp\left(i\frac{\pi}{2\lambda2^k}\hat{q}\hat{\sigma}_y^{(k)}\right) \\
\hat{W}_k &=  \begin{cases}\exp\left( i \frac{\lambda 2^k}{2}\hat{p}\hat{\sigma}_x^{(k)}\right), &  \textrm{if}\, k <N \\
 \exp\left(-i \frac{\lambda 2^k}{2}\hat{p}\hat{\sigma}_x^{(k)}\right), & \textrm{if}\, k  =N\end{cases}
 \label{eq:Rabi}
 \end{split}
\end{align}
where $\hat{q}$ and $\hat{p}$ are the quadrature operators of the CV mode satisfying the commutation relation $[\hat{q},\hat{p}]=i$ and $\sigma_{x}^{(k)}$ and $\sigma_{y}^{(k)}$ are the Pauli-$x$ and $y$ operators of the $k$'th qubit. The interaction parameter $\lambda$ is the only free parameter of the protocol. As we show below, it should be optimized according to the number of qubits and the size of the input state, i.e. the wideness of the support of the input state in phase space. Importantly, a single value of $\lambda$ can be used to encode a wide range of different CV states, meaning that little knowledge of the input CV state is required for the protocol to work.
\begin{figure*}
\centering
\includegraphics{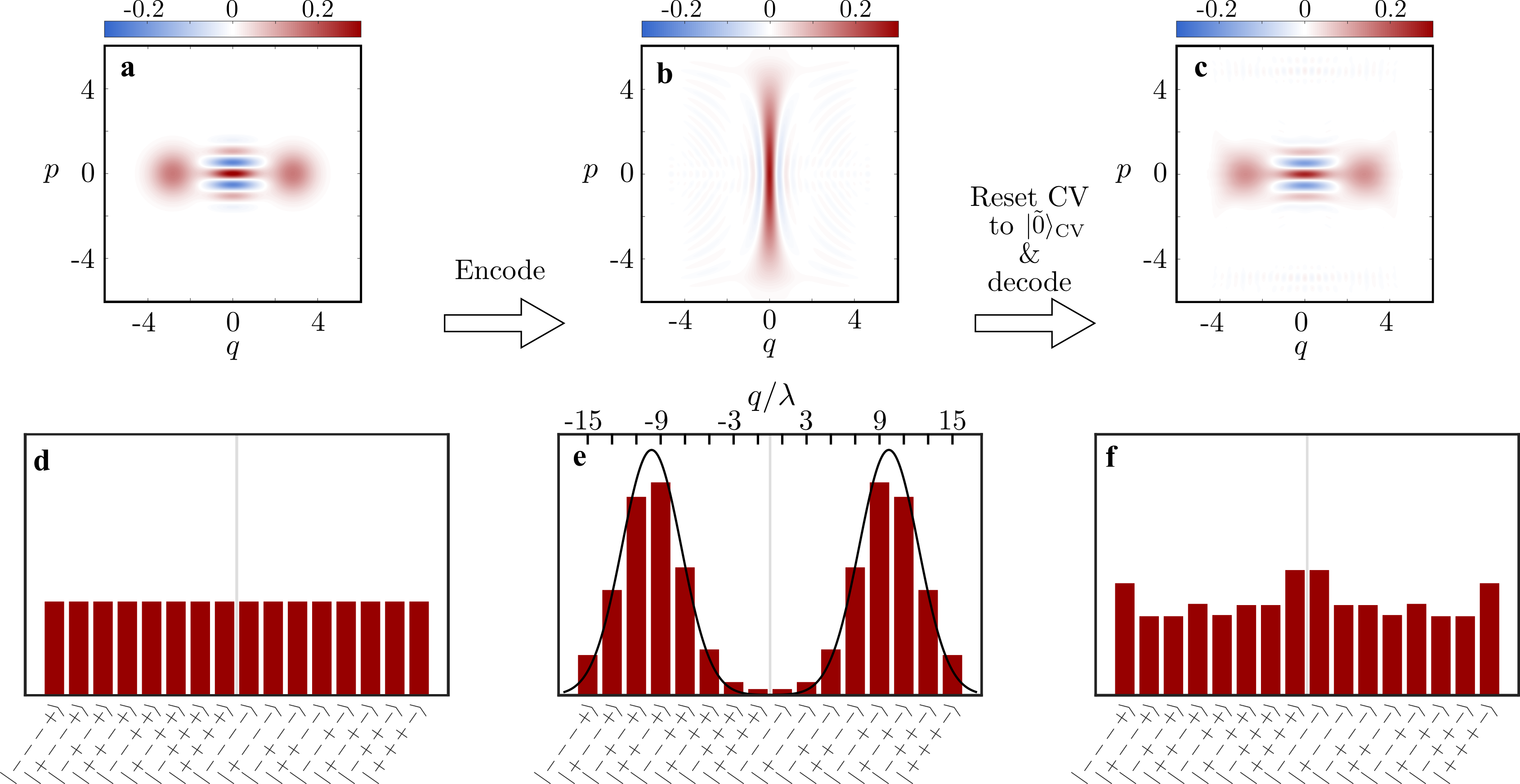}
\caption{Example of encoding and recovery of a CV Schrödinger's cat state, $(e^{-i\sqrt{2}\alpha\hat{p}} + e^{i\sqrt{2}\alpha\hat{p}})\ket{\textrm{vac}}$ with $\alpha = 2$, using $N=4$ qubits and $\lambda = 0.29$. Wigner functions (a,b,c) of the CV mode and probability distributions (d,e,f) of the qubits before encoding (a,d), after encoding (b,e), and after decoding (c,f) with the CV mode completely reset to the state $\ket{\tilde{0}}_\textrm{CV}$ after the encoding. The black curve in (e) shows the $q$ quadrature distribution of the input CV state with the x-axis shown at the top of the figure.}
\label{fig:states}
\bigskip
\includegraphics{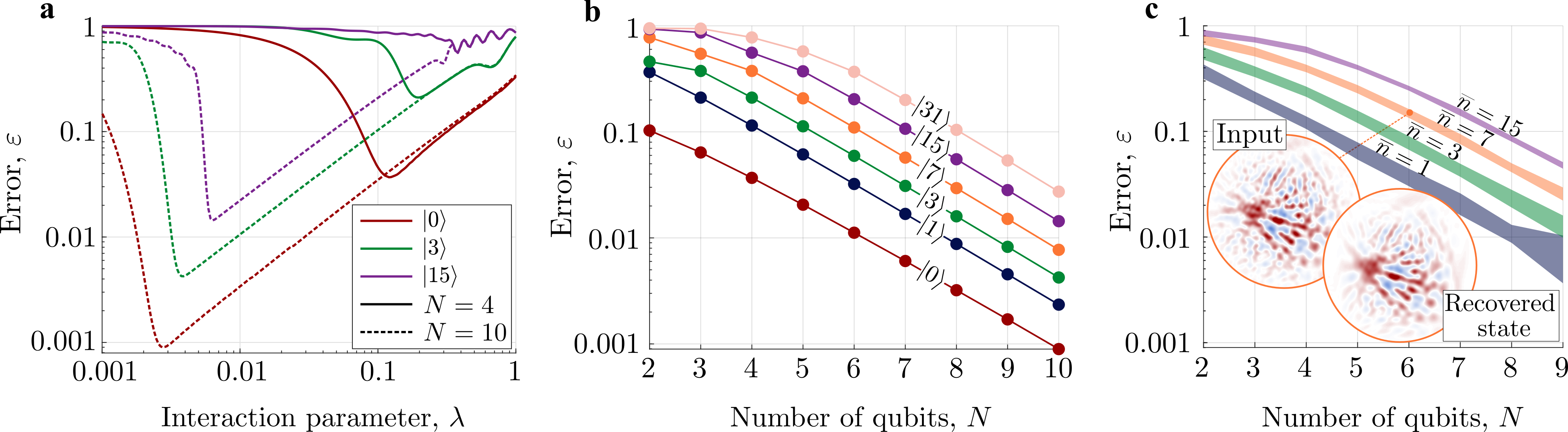}
\caption{(a): Error, $\varepsilon$, as a function of the interaction parameter $\lambda$ for $N=4$ and $10$ qubits with Fock state inputs. (b): Error as a function of qubit number for Fock state inputs using the optimal $\lambda$ for each state. (c): Error as a function of qubit number randomly sampled input states with different fixed mean photon number $\overline{n}$. The shaded areas contains states with $\varepsilon$ within one standard deviation from the mean $\varepsilon$ of the sample. The inserts show the Wigner function of an example input state with $\overline{n}=7$ and the corresponding recovered state using $N=6$ qubits.}
\label{fig:performance}
\end{figure*}
In Appendix B we show that the interaction defined in Eqs.\ \eqref{eq:product} and \eqref{eq:Rabi} achieves the desired unitary operation of equation \eqref{eq:map} for arbitrary states, with $\varepsilon$ decreasing with $N$. The qubit state after the interaction is:
\begin{equation}
    \ket{\Psi}_\textrm{DV} \appropto \sum_\textbf{s} \psi(q_\mathbf{s})\ket{\phi_\textbf{s}} \label{eq:Psi}
\end{equation}
where the sum is over $2^N$ terms, $\ket{\phi_\mathbf{s}}$ form a specific orthonormal basis of the qubit space, $\psi$ is the $q$-quadrature wavefunction of the input CV state and $q_\mathbf{s}$ form an equidistant array of $2^N$ numbers from $-\lambda(2^N-1)$ to $\lambda(2^N-1)$ with spacing $2\lambda$ (see Appendix B for details). Thus the qubit state samples the wave function at $2^N$ discrete points.From this feature we can intuitively understand how we should tune $\lambda$: First, to accurately capture variations in the CV wavefunction, the distance between the samples should be smaller than any large variation of $\psi$, i.e. $2\lambda$ should be sufficiently small. Second, to capture the entire wavefunction the sampling axis should be sufficiently wide, i.e. $\lambda(2^N-1)$ should be large. Satisfying both of these constraints becomes easier for larger $N$, and for fixed $N$ we can expect an optimum $\lambda$ to exist.

The state $\ket{\tilde{0}}_\textrm{CV}$ is given by:
\begin{equation}
    \ket{\tilde{0}}_\textrm{CV} = \frac{1}{\sqrt{2\lambda}}\int dq \sinc\left(\pi \frac{q}{2\lambda}\right)\ket{q}
\end{equation}
where $\sinc(x) = \sin(x)/x$ and $\ket{q}$ denotes a $\hat{q}$ eigenstate, e.g. $\hat{q}\ket{q}=q\ket{q}$. To decode the CV state with the inverse unitary, $\hat{U}^\dagger$, the CV mode should first be prepared in the state $\ket{\tilde{0}}_\textrm{CV}$. Since the encoding protocol approximately leaves the CV mode in state $\ket{\tilde{0}}_\textrm{CV}$, this can be done by applying $\hat{U}$ to an arbitrary CV state, e.g. vacuum or a thermal state, along with ancilliary qubits. In fact, to prepare $\ket{\tilde{0}}_\textrm{CV}$ it suffices to use the same qubit for all $N$ interactions, by resetting the qubit to its ground state after each $\hat{W}\hat{V}$ interaction. Alternatively, $\ket{\tilde{0}}_\textrm{CV}$ can be approximated with fidelity 0.89 by a squeezed vacuum state with squeezing parameter $\textrm{log}(1.12/\lambda)$ (details in Appendix C).
We note that the exact state $\ket{\tilde{0}}_\textrm{CV}$ is in fact unphysical, as it has infinite energy since $\bra{\tilde{0}}\hat{q}^2\ket{\tilde{0}}=\infty$ for all $\lambda$. However, finite energy states, e.g. the state prepared by applying $\hat{U}$ to vacuum, can approximate $\ket{\tilde{0}}_\textrm{CV}$ with high fidelity.

An example of the encoding and recovery of a CV Schrödinger's cat state is shown in Fig.\ \ref{fig:states}. Fig.\ \ref{fig:states}e shows how the input CV wavefunction is directly mapped onto the qubits (with a suitable qubit basis choice). Meanwhile, Fig.\ \ref{fig:states}b shows how the CV mode approximately transforms to the state $\ket{\tilde{0}}_\textrm{CV}$. The state shown in Fig.\ \ref{fig:states}c is the recovered state after the CV mode is completely set to $\ket{\tilde{0}}_\textrm{CV}$ and the qubits are decoded onto the CV mode, i.e. as shown in the circuit of Fig.\ \ref{fig:circuit}a. The small differences between Fig.\ \ref{fig:states}a and c are due to the non-zero $\varepsilon$ arising from the mapping. However, the key features of the CV state, such as the position of the coherent peaks and the central interference pattern with negative values are preserved.

We now numerically demonstrate this result for specific input states. We first consider Fock states, as these represent fundamental quantum basis states, spanning the entire CV mode, with experimentally relevant quantum states typically having main support on low-photon-number Fock states. Fig.\ \ref{fig:performance}a shows how $\varepsilon$ depends on $\lambda$ for $N=4$ and $N=10$ qubits respectively, using Fock states as inputs. For each input we find that there exists an optimum $\lambda$ as expected, and that as we add more qubits, this optimum shifts to smaller values. We also find that, for any fixed $\lambda$, smaller number Fock states are better encoded than larger number Fock states. Thus one setting optimized to encode large states can simultaneously be used to encode smaller states with as good or better performance.

Fig.\ \ref{fig:performance}b shows how $\varepsilon$ depends on the number of qubits for Fock state inputs, choosing the optimum $\lambda$ for each point. We observe a clear exponential decrease in  $\varepsilon$ with increasing number of qubits. Additionally, fixing $\varepsilon$ we find that adding a single qubit allows the storage of approximately twice as large input states, e.g. 4 qubits enable the encoding of $|1\rangle$ with $\varepsilon=0.1$ while 5 qubits allow the encoding of $|3\rangle$ with the same error, 6 qubits can encode $|7\rangle$ and so on. This exponential scaling implies that very large CV states can be encoded using relatively few qubits. 

To demonstrate the versatility of the protocol, Fig.\ \ref{fig:performance}c shows the performance for randomly sampled input states. These states are generated by picking a vector with random complex entries, representing the input state in the Fock-basis. The vector is then filtered with an exponential envelope, damping high Fock-number terms. The strength of the filter is chosen such that a targeted mean photon number $\overline{n}$ is obtained (see Appendix D for details). A typical example of the Wigner function of a resulting random state with $\overline{n}=7$ is shown in the inset of Fig.\ \ref{fig:performance}c. For each $N$ and $\overline{n}$ in Fig.\ \ref{fig:performance}c we calculate $\varepsilon$ for 100 of such random states using a single $\lambda$ chosen to approximately optimize the average $\varepsilon$. The shaded area denotes the states within one standard deviation from the mean $\varepsilon$ of the samples. As with the Fock states, we observe an exponential decrease in $\varepsilon$ with $N$. In addition, we again note that adding a single qubit allows the encoding of states with approximately twice the mean photon number, keeping $\varepsilon$ fixed.

\begin{figure}[t!]
\centering
\includegraphics{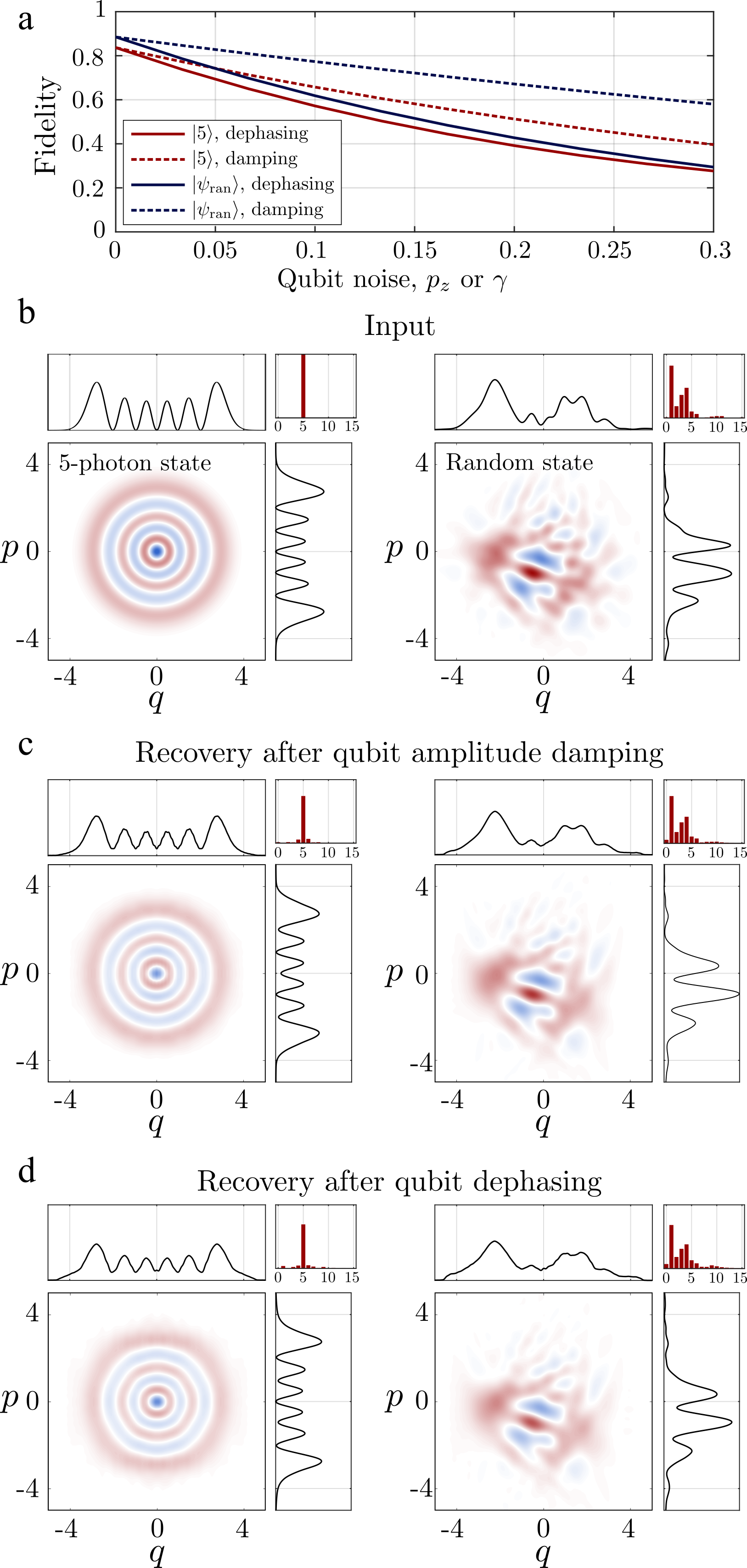}
\caption{(a): Fidelity of recovered states when the qubits undergo dephasing or amplitude damping for an input 5-photon Fock state, and a random state with $\overline{n}=3$ using $\lambda=0.07$ and $N=6$ qubits. (b-d): Wigner functions, quadrature distributions and photon number distributions for the 5-photon Fock state (left) and the $\overline{n}=3$ photon random state (right). (b): Input states. (c): Output when each qubit undergoes dephasing with $p_z=0.05$ (d): Output when each qubit undergoes amplitude damping with $\gamma=0.05$.}
\label{fig:errors}
\end{figure}
Next, we check the stability of our scheme against errors occurring in the qubit system while the state is encoded. In particular, we consider the qubit dephasing channel,
\begin{equation}
\Lambda_z(\rho)= \hat{K}_z^{(1)}\rho(\hat{K}_z^{(1)})^\dagger+\hat{K}_z^{(2)}\rho(\hat{K}_z^{(2)})^\dagger,
\end{equation}
and qubit amplitude damping channel
\begin{equation}
\Lambda_z(\rho)= \hat{K}_\gamma^{(1)}\rho(\hat{K}_z^{(1)})^\dagger+\hat{K}_\gamma^{(2)}\rho(\hat{K}_z^{(2)})^\dagger,
\end{equation}
with Kraus operators:
\begin{align}
    &\hat{K}_z^{(1)} = \sqrt{1-p_z}\hat{I}, && \hat{K}_z^{(2)} = \sqrt{p_z}\hat{\sigma}_z\\
    &\hat{K}_\gamma^{(1)} = \ket{0}\bra{0} + \sqrt{1-\gamma}\ket{1}\bra{1}, && \hat{K}_\gamma^{(2)} = \sqrt{\gamma}\ket{0}\bra{1}
\end{align}
where $\rho$ denotes the qubit density matrix, $p_{z}$ denotes the probability of a single-qubit phase-flip and $\gamma$ denotes the probability of a single qubit decay event. Fig.\ \ref{fig:errors}a shows the fidelity of the recovered state after the CV mode is reset and each qubit have experienced either dephasing or amplitude damping for an input 5-photon Fock state, $\ket{5}$, and a random state with $\overline{n}=3$ average photons, $\ket{\psi_\text{ran}}$ using $N=6$ qubits. As can be expected, the fidelity drops as the qubits experience more noise. However, a single figure of merit, such as the fidelity, is often insufficient to capture the full non-classical aspects of non-Gaussian CV states. Therefore, we also qualitatively analyze the Wigner functions, quadrature distributions and photon distributions of the two selected non-Gaussian trial states. Other input states have shown similar behavior. Fig.\ \ref{fig:errors}c and d shows the recovered states after each qubit has undergone dephasing or amplitude damping with an error probability of $p_z=0.05$ or $\gamma = 0.05$. For both channels we observe a smearing of the q-quadrature distributions while, the p-quadrature distributions remains almost intact compared to the input for both trial states. More importantly, we find that the negative regions of the Wigner functions (highlighted in blue), which are strong indicators of non-classicality, remain non-negligible. Thus even moderate error rates do not have a severe effect on the recovered states. 

In conclusion we have presented a feasible unitary protocol to map arbitrary CV states into a few qubits. This can be realized using only conditional displacements generated by Rabi-type coupling Hamiltonians, which currently are available in trapped-ion systems \cite{fluhmann2018sequential} and superconducting circuits \cite{campagne2020quantum}. The protocol is fully deterministic and requires no measurements or feed-forward. The error rates caused by the finite dimensionality of the qubit subsystem decrease exponentially with the number of qubits. Furthermore, small dephasing or amplitude-damping errors acting on the qubits do not translate into large errors in the protocol. We have focused on encoding arbitrary CV states into qubits, but similar techniques might be used to map arbitrary multi-qubit states into a single CV mode. Such mapping could facilitate multi-qubit operations and hardware-efficient qubit transfers. We leave this as an interesting open direction for future work. 

\section{Acknowledgements}
This  project  was  supported  by  the  Danish  National Research  Foundation  through  the  Center  of  Excellence for Macroscopic Quantum States (bigQ, DNRF0142). RF acknowledges project 21-13265X of the Czech Science Foundation.
\section{References}
\bibliography{References} 

\begin{widetext}
\section{Appendix A: Fidelity of recovered states}
\noindent Here we derive Eq.\ \eqref{eq:fidelity} of the main text. We begin with the definition of the mapping (Eq.\ \eqref{eq:map} of the main text):
\begin{align}
    \hat{U}\ket{\psi}\ket{\textbf{0}} &= \sqrt{1-\varepsilon}\ket{\tilde{0}}\ket{\Psi} + \sqrt{\varepsilon}\ket{\Phi_\varepsilon}.
\end{align}
Resetting the CV mode to state $\ket{\tilde{0}}$ transforms the state into:
\begin{equation}
    \ket{\tilde{0}}\bra{\tilde{0}}\otimes \text{Tr}_\text{CV}\left(\hat{U}\ket{\psi}\bra{\psi}\otimes\ket{\mathbf{0}}\bra{\mathbf{0}}\hat{U}^\dagger\right)=\ket{\tilde{0}}\bra{\tilde{0}}\otimes[(1-\varepsilon)\ket{\Psi}\bra{\Psi} + \varepsilon\rho_\varepsilon]
\end{equation}
where $\rho_\varepsilon=\textrm{Tr}_\textrm{CV}(\ket{\Phi_\varepsilon}\bra{\Phi_\varepsilon})$. Applying $\hat{U}^\dagger$ and calculating the overlap with $\ket{\psi}\ket{\textbf{0}}$ we find:
\begin{align}
    F &= \bra{\psi}\bra{\textbf{0}}\hat{U}^\dagger\left(\ket{\tilde{0}}\bra{\tilde{0}}\otimes[(1-\varepsilon)\ket{\Psi}\bra{\Psi} + \varepsilon\rho_\varepsilon]\right)\hat{U}\ket{\psi}\ket{\textbf{0}}\\
    &= \left(\sqrt{1-\varepsilon}\bra{\tilde{0}}\bra{\Psi} + \sqrt{\varepsilon}\bra{\Phi_\varepsilon}\right)\left(\ket{\tilde{0}}\bra{\tilde{0}}\otimes[(1-\varepsilon)\ket{\Psi}\bra{\Psi} + \varepsilon\rho_\varepsilon]\right)\left(\sqrt{1-\varepsilon}\ket{\tilde{0}}\ket{\Psi} + \sqrt{\varepsilon}\ket{\Phi_\varepsilon}\right) \\
    & = (1-\varepsilon)\bra{\Psi}\left[(1-\varepsilon)\ket{\Psi}\bra{\Psi} + \varepsilon\rho_\varepsilon \right]\ket{\Psi}\\
    & = (1 - \varepsilon)^2 + (1 - \varepsilon)\varepsilon\bra{\Psi}\rho_\varepsilon\ket{\Psi}
\end{align}
where we used $\bra{\tilde{0}}\Phi_\varepsilon\rangle = 0$ and $\bra{\tilde{0}}\tilde{0}\rangle=1$ in the third equality. Since $0\leq \bra{\Psi}\rho_\varepsilon\ket{\Psi}\leq 1$ we have
\begin{equation}
    (1-\varepsilon)^2\leq F \leq (1-\varepsilon)^2 + (1-\varepsilon)\varepsilon = 1 - \varepsilon
\end{equation}
\section{Appendix B: Analytical analysis of the protocol}
\noindent Here we show that the proposed circuit approximately transfers arbitrary CV states onto the qubits. As stated in the main text, the relevant interaction operators are:
\begin{align}
    \hat{V}_k&=\exp\left(iv_k\hat{q}\hat{\sigma}_y^{(k)}\right),\\
    \hat{W}_k&=\begin{cases}\exp\left(iw_k\hat{p}\hat{\sigma}_x^{(k)}\right), &  \text{for $k<N$}, \\ \exp\left(-iw_k\hat{p}\hat{\sigma}_y^{(k)}\right), &  \text{for $k=N$},
    \end{cases}
\end{align}
with
\begin{align}
  v_k &= \frac{\pi}{2\lambda2^k}, \label{eq:u}\\
  w_k &= \frac{\lambda2^k}{2}. \label{eq:v} 
\end{align}
We begin by considering the action of the operators $\hat{V}_k$ and $\hat{W}_k$ on the $\hat{q}$ eigenstates, $\ket{q}$. Using the relations $\exp\left(i\alpha\hat{q}\right)\ket{q} = e^{i\alpha q}\ket{q}$ and $\exp\left(i\alpha\hat{p}\right)\ket{q} = \ket{q - \alpha}$ we get:
\begin{align}
\hat{V}_k\ket{q}\ket{0}_k &= \frac{1}{\sqrt{2}}\left(e^{i v_k q}\ket{q}\ket{i}_k + e^{-i v_k q}\ket{q}\ket{-i}_k\right)\\
& = \frac{e^{i\pi/4}}{2}\bigg[\left(e^{iv_k q} - ie^{-i v_k q}\right)\ket{q}\ket{+}_k + \left(-ie^{iv_k q} + e^{-i v_k q}\right)\ket{q}\ket{-}_k\bigg] \\
&= \cos\left(\frac{\pi}{4} + v_k q\right)\ket{q}\ket{+}_k + \cos\left(\frac{\pi}{4} - v_k q\right)\ket{q}\ket{-}_k,
\end{align}
where $\ket{\pm i}_k= (\ket{0}_k \pm i\ket{1}_k)/\sqrt{2}$ are the $\hat{\sigma}_y^{(k)}$ eigenstates and $\ket{\pm} = (\ket{0}_k \pm \ket{1}_k)/\sqrt{2}$ are the $\hat{\sigma}_x^{(k)}$ eigenstates. Applying $\hat{W}_k$:

\begin{align}
    \hat{W}_k\hat{V}_k\ket{q}\ket{0}_k = \cos\left(\frac{\pi}{4} + v_k q\right)\ket{q - w_k}\ket{+}_k + \cos\left(\frac{\pi}{4} - v_k q\right)\ket{q - (-w_k)}\ket{-}_k \label{eq:VkUk}
\end{align}
Iterating Eq.\ \eqref{eq:VkUk} we get the output for the sequence of operations. For example, after interaction with the first two qubits (with $N>2$) we get:
\begin{align}
    \hat{W}_2\hat{V}_2\hat{W}_1\hat{V}_1\ket{q}\ket{0}_1\ket{0}_2 = &\cos\left(\frac{\pi}{4} + v_1 q\right)\cos\left(\frac{\pi}{4} + v_2 (q - w_1)\right)\ket{q - (w_1 + w_2)}\ket{+}_1\ket{+}_2 \\
    &+ \cos\left(\frac{\pi}{4} - v_1 q\right)\cos\left(\frac{\pi}{4} + v_2 (q - (-w_1))\right)\ket{q - (-w_1 + w_2)}\ket{-}_1\ket{+}_2 \\
    & + \cos\left(\frac{\pi}{4} + v_1 q\right)\cos\left(\frac{\pi}{4} - v_2 (q - w_1)\right)\ket{q - (w_1 - w_2)}\ket{+}_1\ket{-}_2 \\
    & + \cos\left(\frac{\pi}{4} - v_1 q\right)\cos\left(\frac{\pi}{4} - v_2 (q - (-w_1))\right)\ket{q - (-w_1 - w_2)}\ket{-}_1\ket{-}_2
\end{align}
By induction, after interaction with all $n$ qubits we obtain the following expression:
\begin{equation}
    \prod_{k=1}^N\hat{W}_k\hat{V}_k\ket{q}\ket{\textbf{0}} = \sum_\mathbf{s}\prod_{k=1}^N\cos\left[\frac{\pi}{4} + s_k v_k\left(q - \sum_{l=1}^{k-1}s_l w_l\right)\right]\Big|q - \sum_{l=1}^{N-1} s_l w_l + s_N w_N\Big \rangle\ket{\phi_\mathbf{s}}
\end{equation}
 where $\ket{\mathbf{0}}=\ket{0}_1\ket{0}_2...\ket{0}_N$ is the joint qubit ground state and the sum with summation index $\mathbf{s} = (s_1,s_2,...,s_N)$ with $s_k=\pm 1$ is over all $2^N$ combinations of the signs, $s_k$. $\ket{\phi_\mathbf{s}}$ is a qubit product state where mode $k$ is in the $s_k$ eigenstate of $\hat{\sigma}_x^{(k)}$, e.g., $\ket{\phi_{(1,1,-1,-1)}} = \ket{+}_1\ket{+}_2\ket{-}_3\ket{-}_4$ and  $\ket{\phi_{(1,-1,1,1)}} = \ket{+}_1\ket{-}_2\ket{+}_3\ket{+}_4$. By linearity, we can extend the action of the operators to an arbitrary pure state, $\ket{\psi}=\int dq \psi(q)\ket{q}$:
\begin{align}
\prod_{k=1}^N\hat{W}_k\hat{V}_k\ket{\psi}\ket{\textbf{0}} &=  \sum_\mathbf{s}\int dq \psi(q)\prod_{k=1}^N\cos\left[\frac{\pi}{4} + s_k v_k\left(q - \sum_{l=1}^{k-1}s_l w_l\right)\right]\Big|q - \sum_{l=1}^{N-1} s_l w_l + s_N w_N\Big \rangle\ket{\phi_\mathbf{s}}
\intertext{Translating the integration variable, $q \rightarrow q + (\sum_{l=1}^{N-1}s_l w_l - s_N w_N) \equiv q + q_\mathbf{s}$:}
&= \sum_\mathbf{s}\int dq  \psi(q + q_\mathbf{s})\prod_{k=1}^N\cos\left[\frac{\pi}{4} + s_k v_k\left(q - \sum_{l=1}^{k-1}s_l w_l + \sum_{l=1}^{N-1}s_l w_s - s_N w_N\right)\right]\ket{q}\ket{\phi_\mathbf{s}} \\
&=  \sum_\mathbf{s}\int dq \psi(q + q_\mathbf{s})\prod_{k=1}^N\cos\left[\frac{\pi}{4} + s_k v_k\left(q + \sum_{l=k}^{N-1}s_l w_l - s_N w_N\right)\right]\ket{q}\ket{\phi_\mathbf{s}} \\
&=  \sum_\mathbf{s}\int dq \psi(q + q_\mathbf{s})\prod_{k=1}^N\cos\left[\frac{\pi}{4}s_k + v_k q + \sum_{l=k}^{N-1}s_l v_k w_l - s_N v_k w_N\right]\ket{q}\ket{\phi_\mathbf{s}} \\
\intertext{Now consider the sum over $l$ inside the cosine: From our choice of $v_k$ and $w_k$ (Eqs.\ \eqref{eq:u} and \eqref{eq:v}) we get $v_kw_l = 2^{l-k}\pi/4$. Thus when $l\geq k +3 $, $s_lv_kw_l$ is a multiple of $2\pi$, which can be ignored inside the cosine. When $l= k + 2$ we get $s_lv_kw_l = s_{k+2}v_k w_{k+2} = s_{k+2}\pi$, which takes $\cos$ to $-\cos$, regardless of the sign $s_{k+2}$. This term thus contributes an overall $-1$ phase factor which can be ignored. The remaining relevant terms are thus $l=k$ and $l = k + 1$:}
&=  \sum_\mathbf{s}\int dq \psi(q + q_\mathbf{s})\prod_{k=1}^{N-2}\cos\left[\frac{\pi}{2}( s_k + s_{k+1}) + v_k q\right]\cos\left[\frac{\pi}{2}(s_{N-1} - s_{N}) + v_{N-1} q\right]\cos\left[v_N q\right]\ket{q}\ket{\phi_\mathbf{s}} \\
\intertext{The $\pi/2$ terms either add up to $\pm\pi$ if $s_k = s_{k-1}$ or to 0 if $s_k = -s_{k-1}$. Each $\pm\pi$ inside a cosine contributes a $-1$ phase factor, so depending on $\mathbf{s}$ the total phase is either $+1$ or $-1$:}
&= \sum_\mathbf{s}\int dq  (-1)^{\gamma_\mathbf{s}}\psi(q + q_\mathbf{s})\prod_{k=1}^{N}\cos\left(v_k q\right)\ket{q}\ket{\phi_\mathbf{s}}\\
&= \sum_\mathbf{s}\int dq  (-1)^{\gamma_\mathbf{s}}\psi(q + q_\mathbf{s})\prod_{k=1}^{N}\cos\left(\pi\frac{q}{2\lambda 2^k}\right)\ket{q}\ket{\phi_\mathbf{s}},
\end{align}
\begin{figure}
    \centering
    \includegraphics{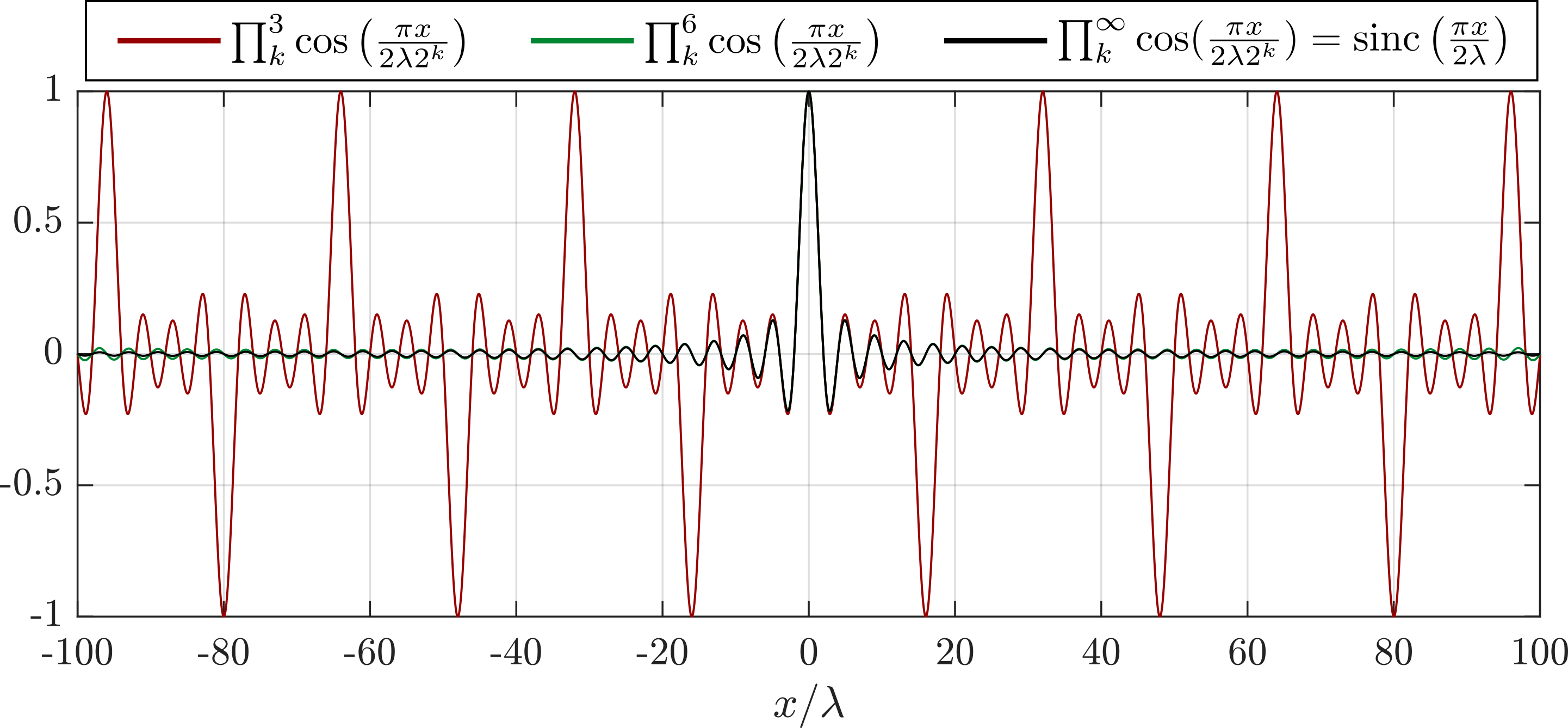}
    \caption{Comparison of functions of the type $\prod_{k=1}^{N}\cos\left(\pi\frac{x}{2\lambda 2^k}\right)$ for $N=3$, $N=6$ and $N=\infty$.}
    \label{fig:Sinc}
\end{figure}
\noindent where $\gamma_\textbf{s} = \sum_{k=1}^{N-2}(s_k + s_{k+1})/2 + (s_{N-1} - s_N)/2$. This phase factor due to $\gamma_\mathbf{s}$ can be absorbed into $\ket{\phi_\mathbf{s}}$ as $(-1)^{\gamma_\mathbf{s}}\ket{\phi_\mathbf{s}}\equiv\ket{\bar{\phi}_\mathbf{s}}$. 

For $N\gg 1$ the term $\prod_{k=1}^{N}\cos\left(\pi\frac{q}{2\lambda 2^k}\right)$ is peaked at $q$ around integer multiples of $\lambda2^{N+1}$ with a peak width on the order of $\lambda$. In particular, for $N=\infty$ we get $\prod_{k=1}^{\infty}\cos\left(\pi\frac{q}{2\lambda 2^k}\right) = \sin(\pi\frac{q}{2\lambda})/(\pi \frac{q}{2\lambda}) \equiv \textrm{sinc}\left(\pi \frac{q}{2\lambda}\right)$, which approaches $\lambda\delta(q/2)$ for $\lambda\rightarrow 0$ where $\delta(x)$ is the Dirac delta function. Fig.\ \ref{fig:Sinc} shows a comparison of $\prod_{k=1}^{N}\cos\left(\pi\frac{x}{2\lambda 2^k}\right)$ for $N=3$, $N=6$ and $N = \infty$.

With this in mind, we make two approximations: First, we approximate $\prod_{k=1}^{N}\cos\left(\pi\frac{q}{2\lambda 2^k}\right) \approx \sinc\left(\pi \frac{q}{2\lambda}\right)$. This approximation fails above $|q|=\lambda 2^{N+1}$, so we should ensure that $\psi(q + q_\mathbf{s})$ goes to $0$ before the approximation fails. Since $q_\mathbf{s} \in [-\lambda(2^N-1); \lambda(2^N-1)]$, we should ensure that $\psi$ has vanishing support beyond $\psi(|2^{N+1}\lambda - \lambda(2^N - 1)|) = \psi(|\lambda(2^N - 1)|)$, which can be satisfied by choosing sufficiently large $N$. 

For the second approximation we assume $\int dq \psi(q + q_\mathbf{s})\sinc(\pi\frac{q}{2\lambda}) \approx \int dq \psi(q_\mathbf{s})\sinc(\pi\frac{q}{2\lambda})$. We thus approximate $\psi(q + q_\mathbf{s})$ with its value at $q=0$, as this is where the sinc function is mainly supported. This approximation requires the variation of $\psi$ to be slower than the variation of the sinc function, i.e. $|\frac{d\psi}{dq}| \ll 1/\lambda$, which is satisfied by choosing $\lambda$ sufficiently small. 

In total we get:
\begin{equation}
\prod_{k=1}^N\hat{W}_k\hat{V}_k\ket{\psi}\ket{\textbf{0}} \approx \int dq \sum_{\mathbf{s}}\psi(q_\mathbf{s})\sinc\left(\pi\frac{q}{2\lambda}\right)\ket{q}\ket{\bar{\phi}_\mathbf{s}} = \int dq \frac{\sinc\left(\pi\frac{q}{2\lambda}\right)}{\sqrt{2\lambda}}\ket{q} \otimes \sum_\mathbf{s}\sqrt{2\lambda}\psi(q_\mathbf{s})\ket{\bar{\phi}_\mathbf{s}}.
\end{equation}
Thus the wavefunction $\psi$ has been transferred from the bosonic mode to the qubits. The factor $\sqrt{2\lambda}$ normalizes the CV mode. We note that our approximations require both large $\lambda2^N$ and small $\lambda$. For finite $N$, there thus exists an optimum $\lambda$ which depends on how broadly $\psi$ is supported in phase-space.

\section{Appendix C: Overlap between $\ket{\tilde{0}}_\textrm{CV}$ and squeezed vacuum}
A squeezed vacuum state with squeezing parameter $r$ is given by:
\begin{equation}
    \ket{S(r)} = \frac{e^{r/2}}{\pi^{1/4}}\int dq \exp\left(\frac{(qe^{r})^2}{2}\right)\ket{q}.
\end{equation}
The fidelity between $\ket{S(r)}$ and $\ket{\tilde{0}}_\textrm{CV}$ is:
\begin{align}
    |\langle S(r)|\tilde{0}\rangle|^2 &= \frac{e^r}{2\lambda\sqrt{\pi}}\left[\int dq \sinc \left(\pi\frac{q}{2\lambda} \right) \exp\left(\frac{(qe^{r})^2}{2}\right) \right]^2\\
    &= \frac{2}{\sqrt{\pi}}\lambda e^{r}\textrm{erf}\left(\frac{\pi}{2\sqrt{2}}\frac{1}{\lambda e^{r}}\right)^2,
\end{align}
where $\textrm{erf}(x)= (2/\sqrt{\pi})\int_{0}^xdz \,e^{-z^2}$ is the error function. This expression is optimized for $\lambda e^{r} \approx 1.12 \Leftrightarrow r \approx \log\left(1.12/\lambda\right)$ for which the fidelity takes the value $0.89$.
\section{Appendix D: Random states}
The random states used in this paper are generated as follows: First, we generate a large number (e.g. 200) of complex numbers, $\{c_0,c_1...c_{200}\}$, with uniformly random amplitudes between 0 and 1 and phases between 0 and $2\pi$. From these we construct an unnormalized CV state in the Fock basis as:
\begin{equation}
    \ket{\psi_\textrm{random}} = \sum_m c_m\ket{m}.
\end{equation}
We then apply an exponential filter to dampen high number Fock terms:
\begin{equation}
    \rightarrow \sum_m e^{-\kappa m}c_m\ket{m},
\end{equation}
where $\kappa$ is tuned such that the resulting state has the desired mean photon number. Lastly, the state is normalized.
\end{widetext}
\end{document}